\documentclass[aps,preprint,amssymb,showpacs,tightenlines]{revtex4}

\usepackage{amsmath}
\usepackage{amsthm}
\usepackage{pstricks,pst-node,pst-plot}
\newpsobject{showgrid}{psgrid}{subgriddiv=1, griddots=5, gridlabels=6pt}
\usepackage{graphicx}

\newcommand{\beq}{\begin{equation}}
\newcommand{\eeq}{\end{equation}}
\newcommand{\beqa}{\begin{eqnarray}}
\newcommand{\eeqa}{\end{eqnarray}}
\newcommand{\beg}{\begin{gather}}
\newcommand{\eeg}{\end{gather}}

\begin{document}

\title{Classical and quantum radiation reaction in conformally flat spacetime}

\author{A.~Higuchi$^1$ and P.~J.~Walker$^2$}

\affiliation{Department of Mathematics, University of York,
Heslington, York YO10 5DD, United Kingdom\\ email:
${}^1$ah28@york.ac.uk, ${}^2$pjw120@york.ac.uk}

\date{February 24, 2009}

\begin{abstract}
We investigate the physics of a charged scalar particle moving in conformally flat spacetime with the conformal factor depending only on time in the framework of quantum electrodynamics (QED).  In particular, we show that the radiation-reaction force derived from QED agrees with the classical counterpart in the limit $\hbar\to 0$ using the fact that to lowest order in $\hbar$ the charged scalar field theory with mass $m$ in conformally flat spacetime with conformal factor $\Omega(t)$, which we call Model B, is equivalent to that in flat spacetime with a time-dependent mass $m\Omega(t)$, which we call Model A, at tree level in this limit.  We also consider the one-loop QED corrections to these two models in the semi-classical approximation.  We find nonzero one-loop corrections to the mass and Maxwell's equations in Model A at order $\hbar^{-1}$.   This does not mean, however, that the corresponding one-loop corrections in Model B are nonzero because the equivalence of these models through a conformal transformation breaks down at one loop.  We find that the one-loop corrections vanish in the limit $\hbar \to 0$ in Model B.
\end{abstract}


\maketitle

\newcommand{\vect}[1]{\mathbf{#1}}
\newcommand{\del}{\nabla}
\newcommand{\Lag}{\mathcal{L}}
\newcommand{\diag}{{\rm diag}}
\newcommand{\eps}{\varepsilon}
\newcommand{\bra}[1]{\left\langle{#1}\right|}
\newcommand{\ket}[1]{\left|{#1}\right\rangle}
\newcommand{\scalprod}[2]{\left\langle{#1}|{#2}\right\rangle}
\newcommand{\A}{\mathcal{A}}
\newcommand{\comm}[2]{\left[{#1},{#2}\right]}
\newcommand{\expect}[2]{\left\langle{#1}\right\rangle_{#2}}
\newcommand{\tendsto}{\rightarrow}
\newcommand{\infinity}{\infty}
\renewcommand{\Re}{{\rm Re}}

\section{Introduction}

The physics of accelerated charged particles has been studied since the late nineteenth century, when Larmor published the formula which now bears his name.  His work calculated the power of radiation emitted by an accelerating charge in a non-relativistic setting, and within a year, Li\'enard produced a version which would turn out to be compatible with special relativity.  Since the radiation carries off energy, there must be a force on the charged particle which causes it to lose energy.  This is the Abraham-Lorentz force~\cite{Abraham,Lorentz}, extended to a special-relativistic setting by Dirac in 1938~\cite{Dirac}. (See Ref.~\cite{Poisson} for a review of the Abraham-Lorentz-Dirac force.)
Some decades after the work of Dirac on the radiation-reaction force DeWitt and Brehme~\cite{DeWittBrehme} published work which gave the analogue of the Abraham-Lorentz-Dirac (ALD) result in general curved spacetime.  Their result was subsequently corrected by Hobbs~\cite{Hobbs}, who added some terms describing the effects due to the geometry of spacetime.

The ALD equation is not without its problems, owing to the presence of a term with third-order time derivative.  As a result, it permits run-away solutions, in which the charged particle accelerates on its own without any external force, and the only solution not showing run-away effects contains the so-called pre-acceleration, which violates causality.  These problems, however, never arise if one treats the ALD force as a perturbative force valid only to first order in the fine-structure constant $\alpha$.  In this approach, called the reduction of order~\cite{Landau}, the quantities appearing in the ALD force, e.g. the third-order time derivative of the position, are replaced by the same quantities {\em for the unperturbed motion}.  Thus, the ALD force is treated as an external time-dependent force, which generates no unphysical solutions.  This procedure is usually justified for a charged body with finite size by noting that the ALD force arises only as an approximate force in the zero-size limit. (See, e.g.~Refs.~\cite{Jackson,Rohrlich} for detailed discussion of unphysical solutions to the ALD equations.)

Since classical electrodynamics is an approximation to {\em quantum} electrodynamics (QED), the ALD force should ultimately be justified in the latter theory.  Several authors have analyzed the radiation-reaction problem in QED~\cite{MS,beilok,FordOc,Oc}.  However, 
these authors studied the ALD force with the charged particle/body treated as a quantum particle/body rather than a quantum field.
Recently Higuchi and Martin showed that the radiation-reaction force on a charged particle treated as a quantum field agrees with the ALD force in the $\hbar\to 0$ limit at first order 
in $\alpha$. (Krivitski\v{\i} and Tsytovich~\cite{Tsyt} also derive the ALD force from QED treating the electron as a quantum field, but their approach is different from that of Higuchi and Martin.)
It was shown first that, if the motion is linear, scalar QED produces the same effects as the Abraham-Lorentz force in the non-relativistic approximation~\cite{Higuchi:2002qc}.  The work essentially compared the position expectation value of the one-particle wave function when the scalar field is coupled to the electromagnetic field with that when the fields are not coupled.  In the limit $\hbar \to 0$ this ``position shift'' was shown to be the same as the corresponding classical shift due to the Abraham-Lorentz force.
This work was extended subsequently to the relativistic case~\cite{HiguchiMartin04,HiguchiMartin05}, and then to non-linear motion~\cite{HiguchiMartin06a,HiguchiMartin06b}.  It has also been extended to spinor QED~\cite{unpublished,Martin}.
(In this approach it is clear that the ALD force should be trusted only at first order in $\alpha$, thus the reduction of order being justified.)

A few years ago, Nomura, Sasaki and Yamamoto~\cite{NoSaYa} showed that radiation from a charged scalar particle in QED in conformally flat spacetime is given by the classical Larmor formula for the corresponding flat-spacetime theory in the limit $\hbar\to 0$.
Their result can be generalized to the case where an external electromagnetic field is present and would suggest that scalar QED ought to recover the classical radiation reaction force on a charged scalar particle under an external electromagnetic field in conformally flat spacetime in the limit $\hbar\to 0$.
In this paper we show that this is indeed the case by using the work of Higuchi and Martin with the assumption that the external electromagnetic field and conformal factor depend only on time.

As part of this investigation we also study the one-loop effect on the charged particle.  It has been suggested that, if a charged particle is accelerated by a time-dependent mass term, there will be a one-loop correction to the mass at order $\hbar^{-1}$ in the semi-classical approximation~\cite{Higuchi:2002qc}. (This is not as alarming as it might sound: note that the fine-structure constant $\alpha=e^2/(4\pi \hbar c)=1/137$ is of order $\hbar^{-1}$.) Since a massive charged scalar field in conformally flat spacetime with a time-dependent conformal factor, which we call Model B, is classically equivalent to a charged scalar field in flat spacetime with a time-dependent mass term, which we call Model A, one might expect that there are one-loop corrections of order $\hbar^{-1}$ also in Model B.  We confirm the existence of one-loop corrections of order $\hbar^{-1}$ in Model A but find that there are no one-loop corrections
in Model B in the limit $\hbar \to 0$.  This difference is due to the breakdown of conformal equivalence of Models A and B at one loop.

The rest of this paper is organized as follows.  In Sec.~\ref{sec2} we transform the classical radiation-reaction formula and the Lagrangian 
for electrodynamics in conformally flat spacetime to those in flat spacetime.  We then show that the classical radiation-reaction force is reproduced at tree level by quantum field theory in the limit $\hbar \to 0$ in the sense that the motion of a suitably defined average position of the charged particle agrees with that given by the classical theory.  
In Sec.~\ref{sec3} we consider 
one-loop corrections in charged scalar field theory with time-dependent mass term (Model A). In Sec.~\ref{sec4} we show that there are no one-loop corrections in charged scalar field theory with the conformal factor depending only on time (Model B) in the limit $\hbar\to 0$.  We make some concluding remarks in Sec.~\ref{sec5}.  We use the metric signature $+---$ and let $c=1$ but keep Planck's constant 
$\hbar$ written explicitly throughout this paper unless stated otherwise.

\section{Radiation reaction at tree level}\label{sec2}

In this section we show that the classical electromagnetic radiation-reaction force in conformally flat spacetime with the conformal factor depending only on time is reproduced by QED in the limit $\hbar \to 0$.

We first transform the radiation-reaction force in confomally flat spacetime to that in flat spacetime.
Let $x^\mu(\tau)$ be the world line, parametrized by the proper time $\tau$, of a classical point particle of mass $m$ and charge $e$ in conformally flat spacetime.  
The $4$-velocity is defined by $u^\mu \equiv dx^\mu/d\tau$ and the acceleration by $a^\mu \equiv u^\alpha \nabla_\alpha u^\mu$. We further define $\dot{a}^\mu \equiv u^\alpha \nabla_\alpha a^\mu$. In conformally flat spacetime the DeWitt-Brehme-Hobbs $4$-force is given by
\begin{equation}
f^{(R)\mu} = \frac{2}{3}\alpha_c \left(\dot{a}^\mu - a^2 u^\mu\right)
+ \frac{1}{3} \alpha_c \left(-{R^\mu}_\nu u^\nu + u^\mu R_{\alpha\beta} u^\alpha u^\beta\right),
\end{equation}
where $a^2 \equiv - a^\mu a_\mu$.
We have defined the classical fine-structure constant by
$\alpha_c \equiv e^2/4\pi$. 
Note that the usual fine-structure constant is $\alpha = \alpha_c/\hbar$.
In a spacetime that is not conformally flat there is an additional term called the tail term, which represents the influence of the electromagnetic field generated by the charge itself in the past on its motion.  Since the electromagnetic field propagates on the light-cone in conformally flat spacetime, the tail term is absent.  If there is an external electromagnetic field, 
$F^{\rm  ex}_{\mu\nu}$, then the equation of motion for this particle is
\begin{equation}
m a^\mu = eF^{{\rm  ex}\,\mu\nu}u_\nu + f^{(R)\mu}. \label{field}
\end{equation}

Let the metric be $g_{\mu\nu}(x) = \Omega^2(x) \eta_{\mu\nu}$, where $\eta_{\mu\nu}$ is the metric on flat spacetime. (We leave the conformal factor to be a general function of spacetime point $x = (\mathbf{x},t)$ for now.)  
Let us define the flat-space proper time by
$\tau_\flat\equiv \Omega^{-1} \tau$, the $4$-velocity in the corresponding flat-space theory by 
$u^\mu_\flat \equiv dx^\mu/d\tau_\flat$ and the flat-space 
acceleration by 
$a^\mu_\flat \equiv u^\alpha\partial_\alpha u^\mu_\flat = d^2x^\mu/d\tau_\flat^2$. (We learned the notation with the subscript ``$\flat$'' from Dieter Brill.)  Let us further define $\dot{a}^\mu_\flat \equiv da^\mu_\flat/d\tau_\flat$. Then
\begin{equation}
f^{(R)\mu} = \Omega^{-3}f^{(R)\mu}_\flat \equiv  \frac{2\alpha_c}{3}\Omega^{-3}\left(\dot{a}^\mu_{\flat}
- a^2_\flat u^\mu_\flat\right).  \label{classical}
\end{equation}
Let us define $M(x)\equiv m\Omega(x)$,  Then 
the field equation (\ref{field}) can be written
\begin{equation}
\frac{d\ }{d\tau_\flat}\left[ M(x) u^\mu_\flat\right] - \eta^{\mu\nu}\partial_\nu M(x)
= e\eta^{\mu\alpha}F^{{\rm  ex}}_{\alpha\beta}u_{\flat}^{\beta} + f^{(R)\mu}_\flat.  \label{flateq}
\end{equation}
Thus, the motion of the charged particle of mass $m$ in spacetime with conformally flat metric $g_{\mu\nu}(x) = \Omega^2(x)\eta_{\mu\nu}$ is the same as that of a charged particle with mass $m\Omega(x)$ in flat spacetime under the influence of the ALD force.  We note for later purposes that Eq.~(\ref{flateq}) without the ALD force $f^{ (R)\mu}_\flat$ can be derived as Hamilton's equations from a Hamiltonian given by
\begin{equation}
H(\mathbf{x},\mathbf{p},t) = \sqrt{[\mathbf{p}-e\mathbf{A}^{\rm  ex}
(\mathbf{x},t)]^2 + M^2(\mathbf{x},t)} + eA^{{\rm  ex}\,0}(\mathbf{x},t), \label{Hamiltonian}
\end{equation}
where we have let $\eta^{\mu\nu}A_{\nu}^{{\rm ex}} =(A^{{\rm ex}\,0},\mathbf{A}^{{\rm ex}})$.

Next we transform the Lagrangian for scalar electrodynamics in conformally flat spacetime to that in flat spacetime.
The Lagrangian for scalar QED in general spacetime with metric $g_{\mu\nu}$ and a background electromagnetic field
$A^{\rm  ex}_\mu$ is
\begin{equation}\label{eq:Lag}
\Lag = \sqrt{-g} \left\{-\frac{1}{4} g^{\mu\alpha} g^{\nu\beta} F_{\mu\nu} F_{\alpha\beta} + g^{\mu\nu}
\left({\cal D}_\mu \phi\right)^*{\cal D}_\nu\phi
- \left[(m/\hbar)^2 - \xi R\right] \phi^*\phi \right\},
\end{equation}
where ${\cal D}_\mu\phi \equiv \left[\partial_\mu + iV_\mu/\hbar + i(e/\hbar)A_\mu\right]\phi$, 
$V_\mu \equiv eA_\mu^{\rm  ex}$, and 
$F_{\mu\nu} \equiv \partial_\mu A_\nu - \partial_\nu A_\mu$.  
Now we assume that the metric is conformally flat, i.e.\ $g_{\mu\nu} = \Omega^2\eta_{\mu\nu}$.  
We introduce a rescaled scalar field $\varphi \equiv \Omega\phi$.
Using the fact that Lagrangians are equivalent under addition of total derivative terms, we find
\begin{equation}
\Lag = -\frac{1}{4} \eta^{\mu\alpha} \eta^{\nu\beta} F_{\mu\nu} F_{\alpha\beta} +  \eta^{\mu\nu} ({\cal D}_\mu \varphi)^* 
{\cal D}_\nu \varphi -\left[M_c^2(x)/\hbar^2\right] \varphi^* \varphi, \label{flatQED}
\end{equation}
where
\begin{equation}
M^2_c(x) = m^2\Omega^2 +(6\xi -1)\hbar^2\left(
\eta^{\mu\nu} \partial_\mu \log\Omega \partial_\nu \log\Omega + \eta^{\mu\nu}  \partial_\mu \partial_\nu \log\Omega\right).
\label{massdiff}
\end{equation}
Note that $M^2_c(x) - M^2(x)= M^2_c(x)- m^2\Omega^2(x)$ is of order $\hbar^2$.  

Now we let the conformal factor $\Omega$ and the external electromagnetic field $V_\mu$ depend only on time $t$.  We also let $\Omega(t)\neq 1$ or $V_\mu(t)\neq 0$ only for 
$-T_1 < t < -T_2$ for some positive constants $T_1$ and $T_2$. 
(Thus, this quantum system is disturbed only for a finite period of time in the past of the $t=0$ hypersurface.)  We also choose the gauge $V_0=0$.  
As a result the background field $V_\mu$ satisfies the Lorenz gauge condition, $\eta^{\mu\nu}\partial_\mu V_\nu = 0$.
We will demonstrate that the motion of the particle in scalar QED with 
Lagrangian in Eq.~(\ref{flatQED}) reproduces the classical motion obeying Eq.~(\ref{flateq}) in the limit $\hbar \to 0$ under these conditions.

Since the quantum field $A_\mu$ in the interaction picture satisfies the free field equation $\partial_\alpha\partial^\alpha A_\mu = 0$ in the Feynman gauge, we can expand it as
\begin{equation}
A_\mu(x) = \int \frac{d^3\mathbf{k}}{2k(2\pi)^3}\left[a_\mu(\mathbf{k})e^{-ik\cdot x} + a_\mu^\dagger(\mathbf{k})
e^{ik\cdot x}\right]
\end{equation}
with $k \equiv \|\mathbf{k}\|$,
where
\begin{equation}
\left[a_\mu(\mathbf{k}),a_\nu^\dagger(\mathbf{k}')\right]= -\eta_{\mu\nu}(2\pi)^32\hbar k\delta^3(\mathbf{k}-\mathbf{k}'),\label{commutators}
\end{equation}
with all other commutators vanishing.
The rescaled scalar field $\varphi$ in the interaction picture 
is expanded in terms of the solutions $\Phi_\mathbf{p}(x)$ and $\overline{\Phi}^*_\mathbf{p}(x)$ to its field equation in the background fields,
\begin{align}
&\left\{ \hbar^2 D_\mu D^\mu + M^2_c(t)\right\}\Phi_\mathbf{p}(x) = 0, \label{fieldE}\\
& D_\mu \equiv \partial_\mu + iV_\mu,
\end{align}
and similarly for $\overline{\Phi}^*_\mathbf{p}(x)$, 
such that $\Phi_\mathbf{p}(x) = e^{-ip\cdot x/\hbar}$ and $\overline{\Phi}_\mathbf{p}^*(x)=e^{ip\cdot x/\hbar}$, $p_0 = \sqrt{\mathbf{p}^2+m^2}$, 
for $t > -T_2$.  (The background 
field $V_\mu$ is regarded to be of zeroth order in $e$.) We note that the particle-creation effect is non-perturbative in $\hbar$, i.e. it does not occur at any finite order in $\hbar$, provided that the background fields are smooth, as we assume here.
Hence $\Phi_\mathbf{p}(x) = e^{-ip\cdot x/\hbar + i\delta}$ for some real number $\delta$ also for $t < - T_1$ to all orders in $\hbar$ in the WKB approximation, and similarly for $\overline{\Phi}_\mathbf{p}(x)$.  Then, we can 
expand the scalar field $\varphi$ as
\begin{equation}
\varphi(x) =  \hbar \int\frac{d^3\mathbf{p}}{2p_0(2\pi\hbar)^3}
\left[A(\mathbf{p})\Phi_\mathbf{p}(x) + B^\dagger(\mathbf{p})\overline{\Phi}_\mathbf{p}^*(x)\right],
\end{equation}
where
\begin{equation}
\left[A(\mathbf{p}),A^\dagger(\mathbf{p}')\right]
= \left[B(\mathbf{p}),B^\dagger(\mathbf{p}')\right] = 2p_0(2\pi\hbar)^3\delta^3(\mathbf{p}-\mathbf{p}'),
\end{equation}
with all other commutators vanishing.

To lowest nontrivial order in perturbation theory in the interaction picture, the initial one-particle state 
$\ket{\mathbf{p}} = A^\dagger(\mathbf{p})|0\rangle$ 
evolves to either itself or a state with one photon and one charged particle. Thus,
\begin{equation}
\ket{\mathbf{p}}
\rightarrow \left[1+\frac{i}{\hbar}{\cal F}(\mathbf{p})\right]\ket{\mathbf{p}}
+ \frac{i}{\hbar}\int\frac{d^3\mathbf{k}}{(2\pi)^32k}
{\cal A}^\mu(\mathbf{p},\mathbf{k})a_\mu^\dagger(\mathbf{k})\ket{\mathbf{P}}, 
\label{evolve}
\end{equation}
where $\mathbf{P} = \mathbf{p}-\hbar\mathbf{k}$ by momentum conservation.  We call ${\cal F}(\mathbf{p})$ the forward-scattering amplitude and ${\cal A}^\mu(\mathbf{p},\mathbf{k})$ the one-photon-emission amplitude. 
Then a normalized initial wave-packet state of a charged particle,
\begin{equation}
\ket{I}	= \int \frac{d^3 p}{\sqrt{2p_0}(2\pi\hbar)^3} f(\vect{p})\ket{\mathbf{p}},
\end{equation}
where
\begin{equation}
\int \frac{d^3\mathbf{p}}{(2\pi\hbar)^3}|f(\mathbf{p})|^2 = 1,
\end{equation}
evolves to a final state $|F\rangle$ obtained by replacing the states 
$\ket{\mathbf{p}}$ in $|I\rangle$ by the right-hand side of Eq.~(\ref{evolve}).

Following Refs.~\cite{HiguchiMartin04,HiguchiMartin05,HiguchiMartin06a,HiguchiMartin06b}, we define the average position of the particle at time 
$t > -T_2$ for any state $\ket{\psi}$ with one scalar particle and any number of photons by
\begin{equation}
\langle x^i\rangle_\psi(t) \equiv \int d^3\mathbf{x}\, x^i \langle \psi|\,\rho(\mathbf{x},t)\,|\psi\rangle,
\end{equation}
where $\rho$ is the charge density operator given by
\begin{equation}
\rho(\mathbf{x},t)\equiv \frac{i}{\hbar}:\varphi^\dagger\frac{\partial\varphi}{\partial t}-\frac{\partial\varphi^\dagger}{\partial t}\varphi:,
\end{equation}
with $\varphi$ being now a field operator and $:\cdots:$ denoting normal ordering.  We compare this position expectation value in the final state $|F\rangle$ with that in the initial state $|I\rangle$, which is the position expectation value with $e=0$.  Thus, the change in the position expectation value due to radiation reaction at $t=0$, which we call the position shift, is
\begin{equation}
\delta x^i \equiv \langle x^i \rangle_F(0) - \langle x^i\rangle_I(0).
\end{equation}
It was shown in Refs.~\cite{HiguchiMartin04,HiguchiMartin05,HiguchiMartin06a} that this position shift is given in the limit $\hbar\to 0$ and in the limit where the wave packet is sharply peaked in the momentum space by
\begin{equation}\label{eq:qpe}
\delta x^i = \delta_{\rm tree}x^i + \delta_{\rm loop}x^i,
\end{equation}
where
\beqa
\delta_{\rm tree}x^i & = &  -\frac{i}{2} \int \frac{d^3 k}{2k(2\pi)^3} \A^{\mu*}(\vect{p},\vect{k}) \overleftrightarrow{\partial}_{\!\!p_i} \A_\mu(\vect{p},\vect{k}),\label{treePS}\\
\delta_{\rm loop}x^i & = & - \partial_{p_i} \Re \mathcal{F}(\vect{p}). \label{loopPS}
\eeqa
Now, the corresponding position shift in the classical theory can be calculated using the equation of motion given by  Eq.~(\ref{flateq}) with 
$M(t) = m\Omega(t)$.  
It was shown in Refs.~\cite{HiguchiMartin04,HiguchiMartin05,HiguchiMartin06a} 
that in the limit $\hbar \to 0$ the position shift $\delta x^i$ given by Eq.~(\ref{eq:qpe}) agrees with the classical counterpart due to the ALD force if $M(t)=m={\rm constant}$.  Our main  aim in this paper is to demonstrate this agreement for general time-dependent mass arising as a result of a conformal transformation of a charged scalar field 
theory with time-dependent conformal factor.
In this section we show that the tree-level contribution, $\delta_{\rm tree}x^i$, agrees with the classical position shift.  It will be shown in Sec.~\ref{sec5} that $\delta_{\rm loop}x^i = 0$.  These two results will imply that the quantum position shift agrees with the classical one in the limit
$\hbar\to 0$.

Let us now show that $\delta_{\rm tree}x^i$ reproduces the classical position shift.
First consider the electromagnetic field coupled to a classical external 
current $j^\mu$ with the corresponding Lagrangian,
\beq
{\cal L}' = \sqrt{-g}\left( - \frac{1}{4}F_{\mu\nu}F^{\mu\nu} - eA_\mu j^\mu\right).
\eeq
If the classical current is that of a point charge,
\beq
j^0(\mathbf{x},t) = \delta^3(\mathbf{x}-\mathbf{x}(t)),\,\,\,
j^i(\mathbf{x},t) = \frac{dx^i(t)}{dt}\delta^3(\mathbf{x} - \mathbf{x}(t)),
\eeq
where $\mathbf{x}(t)$ is the position of the particle at time $t$, then the emission amplitude in the Feynman gauge is given by 
\begin{eqnarray}
{\cal A}_{\rm cl}^{\mu}(\mathbf{k}) & \equiv &
- e\int d^4 x \langle 0|a^\mu(\mathbf{k})A_\alpha(x)|0\rangle j^\alpha(x) \nonumber \\
& = & - e\int_{-\infty}^{\infty} d\tau \frac{dx^\mu}{d\tau}e^{ik\cdot x}. \label{classcurr}
\end{eqnarray}
(One needs to cut off the $\tau$ integration smoothly to make this amplitude well defined and remove the cutoff at the end.)  It was shown in Refs.~\cite{HiguchiMartin06a,HiguchiMartin06b} that, if the one-photon-emission amplitude ${\cal A}^\mu(\mathbf{p},\mathbf{k})$ defined by
Eq.~(\ref{evolve}) equals the emission amplitude ${\cal A}^\mu_{\rm cl}(\mathbf{k})$ due to the classical point charge $e$ with the final momentum $\mathbf{p}$ in the same background field, then in the limit $\hbar \to 0$ the tree-level position shift $\delta_{\rm tree}x^i$ given by 
Eq.~(\ref{treePS}) equals the position shift due to the classical ALD force $f^{(R)\,\mu}_\flat$.  A further condition needed in the derivation of this result is that the system with $e=0$ should be a Hamiltonian system.  This condition is satisfied
as we have seen [see 
Eq.~(\ref{Hamiltonian})]. We now show that 
${\cal A}^\mu(\mathbf{p},\mathbf{k})$ equals the classical emission amplitude 
${\cal A}_{\rm cl}^\mu(\mathbf{k})$ for the point charge with final momentum $\mathbf{p}$, and this is sufficient for concluding that $\delta_{\rm tree}x^i$ equals the classical position shift.
(The derivation is almost identical to the case with time-independent mass.) 

The one-particle wave function $\Phi_\mathbf{p}(x)$ satisfying Eq.~(\ref{fieldE}) is given in the WKB approximation as
\beqa
\Phi_\mathbf{p}(x) &  = & \sqrt{p_0}\,\phi_\mathbf{p}(t)e^{i\mathbf{p}\cdot\mathbf{x}/\hbar},\label{Phi2}\\
\phi_\mathbf{p}(t) & = & \frac{1}{\sqrt{\sigma_\mathbf{p}(t)}}\exp\left[-\frac{i}{\hbar}\int_0^t
\sigma_\mathbf{p}(t')dt'\right]\psi_\mathbf{p}(t),  \label{Phi}
\eeqa
where $p_0\equiv \sqrt{\mathbf{p}^2 + m^2}$ and $\sigma_\mathbf{p}(t) \equiv \sqrt{[\mathbf{p}-\mathbf{V}(t)]^2 + M^2_c(t)}$.
The function $\psi_\mathbf{p}(t) = 1 + O(\hbar)$ contains the terms of higher order in $\hbar$.
If $x^\mu(t)$ (with $x^0=t$) is the world line of the classical charged particle of mass $M_c(t)$ in the background field $\mathbf{V}$ 
passing through the spacetime origin with momentum $\mathbf{p}$, then we find
\beqa
\sigma_\mathbf{p}(t) & = & M_c(t)\frac{dt}{d\tau},\label{energy}\\
\tilde{\mathbf{p}}(t) & \equiv & \mathbf{p} -\mathbf{V}(t) = M_c(t)\frac{d\mathbf{x}}{d\tau},\label{momentum}
\eeqa
where $\tau$ is the proper time along the world line, by Hamilton's equations from the Hamiltonian (\ref{Hamiltonian}) with $M^2(x)$ replaced by $M_c^2(t)$ and with $eA_\mu^{\rm ext} = V_\mu(t)$.
Now, let
\beq
J^\mu(x) \equiv
 \frac{i}{\hbar}:\varphi^\dagger(x)D^\mu \varphi(x) - [D^\mu\varphi^\dagger(x)]\varphi(x):.
\label{currentJ}
\eeq
Then 
\beq\label{eq:J^mu}
{\cal A}^\mu(\mathbf{p},\mathbf{k})
= - e\int \frac{d^3\mathbf{P}}{2P_0(2\pi\hbar)^3}
\int d^4 x \langle \mathbf{P}|J^\mu(x)|\mathbf{p}\rangle e^{-i\mathbf{k}\cdot\mathbf{x}} e^{ik t}.
\eeq
We find to lowest nontrivial order in $\hbar$
\beqa
\int d^3\mathbf{x}\,\langle \mathbf{P}|J^\mu(x)|\mathbf{p}\rangle e^{-i\mathbf{k}\cdot\mathbf{x}}
& = & \frac{2\tilde{p}^\mu p_0}{\sigma_\mathbf{p}}\exp\left(i \int_0^t
\frac{\sigma_{\mathbf{P}}(t')-\sigma_\mathbf{p}(t')}{\hbar}\,dt'\right)(2\pi\hbar)^3
\delta^3(\mathbf{P}-\mathbf{p}-\hbar\mathbf{k})\nonumber \\
& = & \frac{2\tilde{p}^\mu p_0}{\sigma_\mathbf{p}} \exp \left(-i\int_0^t \frac{\tilde{\mathbf{p}}\cdot\mathbf{k}}{\sigma_\mathbf{p}(t')}dt'\right)
(2\pi\hbar)^3\delta^3(\mathbf{P}-\mathbf{p}-\hbar\mathbf{k}),
\eeqa
where $\tilde{p}^\mu \equiv (\sigma_\mathbf{p}, \tilde{\mathbf{p}})$.  Upon integration over $\mathbf{P}$ we find in the limit $\hbar\to 0$
\beq
{\cal A}^\mu(\mathbf{p},\mathbf{k})
= - e\int dt \frac{\tilde{p}^\mu}{\sigma_\mathbf{p}} \exp\left(-i\int_0^t
\frac{\tilde{\mathbf{p}}}{\sigma_\mathbf{p}}dt\cdot\mathbf{k}\right)e^{ikt}. \label{intermed}
\eeq
{}From Eqs.~(\ref{energy}) and (\ref{momentum}) we have
%Now the classical motion is given by the following Hamiltonian:
%\beq
%H(\mathbf{x},\mathbf{p},t) = \sqrt{(\mathbf{p}-\mathbf{V})^2 + M^2(t)}.
%\eeq
%(This can be extended to the case were the background electromagnetic field and the mass are both spacetime dependent.)  Then
\beq
\frac{\tilde{\mathbf{p}}}{\sigma_\mathbf{p}(t)}=\frac{d\mathbf{x}}{dt}.
\eeq
By using this formula in Eq.~(\ref{intermed}) we find
\beq
{\cal A}^\mu(\mathbf{p},\mathbf{k}) = - e\int d\tau \frac{dx^\mu}{d\tau}e^{ik\cdot x}.
\eeq
The right-hand side is ${\cal A}_{\rm cl}^\mu(\mathbf{k})$ given by Eq.~(\ref{classcurr}) 
for a point charge whose motion is derived from 
the Hamiltonian (\ref{Hamiltonian}) with $M(t)$ replaced by $M_c(t)$.  Since the difference between these
two masses is of order $\hbar^2$, the one-photon-emission amplitude ${\cal A}^\mu(\mathbf{p},\mathbf{k})$ is equal to the emission amplitude for the corresponding classical particle in the limit $\hbar\to 0$.  Hence, we conclude that the position shift $\delta_{\rm tree}x^i$ is identical to the corresponding classical position shift in the limit $\hbar\to 0$.

\section{One-loop corrections in Model A}\label{sec3}
%\noinden
%{\bf One-loop for time-dependent mass term case}
In this section we calculate the one-loop corrections for the charged scalar field with a time-dependent mass $M(t)$, which is not assumed to arise as a result of conformal transformation, in an external electromagnetic potential $\mathbf{V}(t)$. 
%These one-loop corrections contribute to the forward-scattering amplitude ${\cal F}(\mathbf{p})$, which in turn contributes to %the position shift.  
It was argued in Ref.~\cite{Higuchi:2002qc} that there is a logarithmic correction to the mass that is of order $\hbar^{-1}$.  We confirm this assertion in this section. We also show that the relation between the external current and the electromagnetic field it generates is modified.

Let us assume that an external current $J_{\rm C}^\mu$ generates the background field $V_\mu$ at tree level, i.e.
\beq
\partial_\nu(\partial^\nu V^\mu - \partial^\mu V^\nu) = e^2 J_{\rm C}^\mu.
\eeq
We write the Lagrangian in terms of the renormalized fields and coupling constant as
\beqa
{\cal L} & = & -\frac{Z_3}{4}\tilde{F}_{\mu\nu}\tilde{F}^{\mu\nu}
+ Z_2({\cal D}_\mu \varphi)^\dagger {\cal D}^\mu\varphi -\left[M^2(t)/\hbar^2\right] \varphi^\dagger\varphi\nonumber\\
&& + \left[\delta M^2(t)/\hbar^2\right]\varphi^\dagger\varphi - e\left(J_{{\rm C}}^\mu + \Delta J^\mu\right)
A_\mu, \label{Lagden}
\eeqa
where we have used the Ward identity $Z_1=Z_2$ (see, e.g.\ Ref.~\cite{Itzykson}).  
We use the dimensional regularization~\cite{tHooft}, which allows multiplicative mass renormalization, 
$\delta M^2(t) \propto M^2(t)$, at one loop.
The field strength $\tilde{F}_{\mu\nu}$ is given in terms of the total electromagnetic field 
$\tilde{A}_\mu= e^{-1}V_\mu + A_\mu$ as $\tilde{F}_{\mu\nu} = \partial_\mu \tilde{A}_\nu - \partial_\nu \tilde{A}_\mu$.
The covariant derivative is given by
${\cal D}_\mu = \partial_\mu + i(e/\hbar)\tilde{A}_\mu$.
The current $\Delta J^\mu$ needs to be added in the Lagrangian
to keep the background field $V_\mu$ at one-loop order.
Maxwell's equations at one loop are
%The background field $V^\mu$ is determined by the equation
\beq
Z_3 \partial_\nu (\partial^\nu V^\mu - \partial^\mu V^\nu) = e^2 (J_{{\rm C}}^\mu + \Delta J^\mu + J_{{\rm Q}}^\mu),
\label{one-loopeq}
\eeq
where $J_{{\rm Q}}^\mu = \langle 0|J^\mu|0\rangle$, which we call the vacuum current. [The operator $J^\mu$ is given by Eq. (\ref{eq:J^mu}).] This equation can be written
\beq
e^2\Delta J^\mu  =  e^2J_{\rm Q}^{\mu} -(Z_3-1)\partial_\nu(\partial^\nu V^\mu - \partial^\mu V^\nu).
\label{Z3-1}
\eeq
%where $J_{{\rm Q}}^\mu \equiv \langle 0|J^\mu|0\rangle$ 
%is the quantum correction to the current discussed later in this section. 
The tree-level current $J_{\rm C}^\mu$ needs to be supplemented by $\Delta J^\mu$ in order to produce the external field $V^\mu$ at one-loop order.

By rewriting the total electromagnetic field as
$e\tilde{A}_\mu = V_\mu + eA_\mu$,
substituting it in the Lagrangian (\ref{Lagden}) and dropping total-derivative and field-independent terms, we find
\beqa
{\cal L} & = & - \frac{Z_3}{4}F_{\mu\nu}F^{\mu\nu}
+ Z_2({\cal D}_\mu \varphi)^\dagger {\cal D}^\mu\varphi - \left[M^2(t)/\hbar^2\right]\varphi^\dagger\varphi\nonumber \\
&& + \left[\delta M^2(t)/\hbar^2\right]\varphi^\dagger\varphi
+ eA_\mu\Delta J^\mu.
\eeqa
\subsection{The vacuum current}

In this subsection we show that the current $\Delta J^\mu$ given by Eq.~(\ref{Z3-1}) is nonzero at order $\hbar^{-1}$ in the semi-classical approximation, i.e.\ that Maxwell's equations, which relates the external current to the external electric field, are altered at this order.
%field equation (\ref{one-loopeq}) that determines the classical background field at one loop is different from the tree-level equation,
%\beq
%\partial_\nu(\partial^\nu V^\mu - \partial^\mu V^\nu) = e^2 J_{\rm C}^\mu.
%\eeq
%Since we assume that the field $V^\mu$ has only space components and that it depends only on time, we can write this equation as
%\beq
%\ddot{\mathbf{V}} = e^2 \mathbf{J}_{\rm C}.
%\eeq
%The corresponding equation at one loop can be written
%\beq
%
%\ddot{\mathbf{V}} = e^2 \mathbf{J}_{\rm C} + e^2\mathbf{J}_{\rm Q} - (Z_3-1)\ddot{\mathbf{V}}.  \label{effective}
%\eeq
%We show that $e^2\mathbf{J}_{\rm Q} \neq (Z_3-1)\ddot{\mathbf{V}}$ and, hence, the field equation for $V^\mu$ is modified at %one loop.
We first calculate the vacuum current 
$J_{\rm Q}^\mu = \langle 0|J^\mu|0\rangle$. Under the assumption that $V^0(t)=0$ and that $\mathbf{V}$ depends only on $t$, one can readily show that $J_{\rm Q}^0 = 0$.  This result is physically reasonable because the electric field $e\mathbf{E}(t)= \dot{\mathbf{V}}(t)$ is homogeneous and cannot create an inhomogeneous charge distribution.  The space components of 
$J_{\rm Q}^\mu$ can be written 
\beq
\mathbf{J}_{\rm Q} = \int \frac{d^3\mathbf{p}}{\sigma_\mathbf{p}(t)(2\pi\hbar)^3}
\tilde{\mathbf{p}}|\psi_\mathbf{p}(t)|^2.  \label{qcorr}
\eeq
We evaluate $\mathbf{J}_{\rm Q}$ 
up to terms that vanish as $\hbar \to 0$.  To this end we need to find $|\psi_\mathbf{p}(t)|^2$, 
which turn out to have no terms odd in $\hbar$, to order $\hbar^2$.

Let us define
\beq
\Sigma(t) \equiv \sigma_\mathbf{p}^2 = \tilde{\mathbf{p}}^2 + M^2(t).
\eeq
Then by substituting the definition of $\psi_\mathbf{p}(t)$, 
Eqs.~(\ref{Phi2})-(\ref{Phi}), in Eq.~(\ref{fieldE}) with $M^2_c(t)$ replaced by $M^2(t)$ we have
%\beq
%\left[ \hbar^2\frac{d^2\ }{dt^2}
%%+ \Sigma(t)\right]\left\{
%\Sigma^{-1/4}(t) \exp\left[-\frac{i}{\hbar}
%\int_0^t \Sigma^{1/2}(t')dt'\right]\psi_\mathbf{p}(t)\right\} = 0.
%\eeq
%This can be written as
\beqa
\frac{1}{i\hbar}\Sigma^{1/2}\frac{d\ }{dt}\log\psi_\mathbf{p}
& = & \frac{1}{4}\frac{\dot{\Sigma}}{\Sigma}
\frac{d\ }{dt}\log\psi_\mathbf{p} - \frac{1}{2}\frac{d^2\ }{dt^2}\log\psi_\mathbf{p} - \frac{1}{2}\left(
\frac{d\ }{dt}\log\psi_\mathbf{p}\right)^2 \nonumber \\
&& + \frac{1}{8}\frac{\ddot{\Sigma}}{\Sigma} - \frac{5}{32}\frac{\dot{\Sigma}^2}{\Sigma^2}.  \label{WKBeq}
\eeqa
We let
\beq
\psi_\mathbf{p} = \exp\left[ i\hbar\psi_\mathbf{p}^{(1)} + (i\hbar)^2\psi_\mathbf{p}^{(2)} + \cdots\right].
\eeq
Then 
\beq
|\psi_\mathbf{p}|^2  
%&= & \exp \left[ - 2\hbar^2\psi_\mathbf{p}^{(2)}+\cdots \right]\nonumber \\
 =  1 - 2\hbar^2\psi_\mathbf{p}^{(2)} + O(\hbar^4).  \label{psisquare}
\eeq
Eq.~(\ref{WKBeq}) can be used to find $\psi_\mathbf{p}^{(n)}$, $n=1,2,\ldots$, recursively.  Thus, we find
%We find
%\beqa
%\dot{\psi}_\mathbf{p}^{(1)} = \frac{1}{8}\frac{\ddot{\Sigma}}{\Sigma^{3/2}} - \frac{5}{32}\frac{\dot{\Sigma}^2}{\Sigma^{5/2}}.
%\eeqa
%Next we find
%\beqa
%\dot{\psi}_\mathbf{p}^{(2)} & = & \frac{\dot{\Sigma}}{4\Sigma^{3/2}}\dot{\psi}_\mathbf{p}^{(1)}
%- \frac{1}{2\Sigma^{1/2}}\dot{\psi}_\mathbf{p}^{(1)}\nonumber \\
%& = & - \frac{1}{2}\frac{d\ }{dt}\left[ \frac{\dot{\psi}_\mathbf{p}^{(1)}}{\Sigma^{1/2}}\right].
%\eeqa
%Hence
\beq
\psi_\mathbf{p}^{(2)} = \frac{5\dot{\Sigma}^2}{64\Sigma^3} -
\frac{\ddot{\Sigma}}{16\Sigma^2}. \label{psi2}
\eeq
%Now,
%\beqa
%\dot{\Sigma} & = & - 2\tilde{\mathbf{p}}\cdot\dot{\mathbf{V}} + \frac{d\ }{dt}M^2(t),\\
%\ddot{\Sigma} & = & - 2\tilde{\mathbf{p}}\cdot\ddot{\mathbf{V}}
%+ \|\dot{\mathbf{V}}\|^2 + \frac{d^2\ }{dt^2}M^2(t).
%\eeqa
Substituting Eq.~(\ref{psisquare}), with $\psi_\mathbf{p}^{(2)}(t)$ given by this formula, in Eq.~(\ref{qcorr}) 
we find
%only the terms odd in $\tilde{\mathbf{p}}$ give nonzero results.  Thus, we have
\beq
J^i = J_{Q1}^i + J_{Q2}^i,
\eeq
where
\beqa
e^2J_{\rm Q1}^i & = & \frac{5e^2}{24\hbar}\dot{V}^i\left(\frac{d\ }{dt}M^2(t)\right)
\int\frac{d^3\tilde{\mathbf{p}}}{(2\pi)^3}
\frac{\tilde{\mathbf{p}}^2}{\sigma_\mathbf{p}^7}, \label{Jvac1} \\
e^2J_{\rm Q2}^i & = & - e^2\mu^{4-D}\frac{\ddot{V}^j}{4\hbar}
\int \frac{d^{D-1}\tilde{\mathbf{p}}}{(2\pi)^{D-1}}\frac{\tilde{p}^i\tilde{p}^j}{\sigma_\mathbf{p}^5}.  \label{Jvac}
\eeqa
We have dimensionally regularized the integral over $3$-momentum in $J_{\rm Q2}^i$ with
$D= 4 - 2\varepsilon$, introducing the renormalization scale $\mu$.
One can readily evaluate these integrals with the following results:
%\beq
%\int\frac{d^3\tilde{\mathbf{p}}}{(2\pi)^3}\frac{\tilde{\mathbf{p}}^2}{\sigma_\mathbf{p}^7}
%= \frac{1}{2\pi^2}\int_0^\infty \frac{p^4dp}{(p^2+M^2(t))^{7/2}}.
%\eeq
%This integral can readily be performed by letting $p=M(t)\tan\theta$.  The result is
%\beq
%{\rm LHS} = \frac{1}{5M^2(t)}.
%\eeq
%Then the first term in Eq.~(\ref{Jvac}) is
\beqa
e^2\mathbf{J}_{\rm Q1} & = & \frac{e^2}{48\pi^2\hbar}\left(\frac{d\ }{dt}\log M^2(t)\right)\dot{\mathbf{V}},\label{Q1}\\
%\eeq
%In deriving the second term in Eq.~(\ref{Jvac}) we set
%\beq
%\int d^3\tilde{\mathbf{p}}\tilde{p}^i \tilde{p}^j f(\tilde{\mathbf{p}}^2) = \frac{\delta^{ij}}{3}
%\int d^3\tilde{\mathbf{p}}\tilde{\mathbf{p}}^2f(\tilde{\mathbf{p}}^2).
%\eeq
%Since the second term in Eq.~(\ref{Jvac}) is ultraviolet divergent, we let $3\to d-1 = 3-2\epsilon$.  Then the $3$ in the %denominator here should be replaced by $d-1$, where $d=4-2\epsilon$.  Thus, the integral for the second term in %Eq.~(\ref{Jvac}) should be made
%\beq
%\int \frac{d^3\tilde{\mathbf{p}}}{(2\pi)^3}\frac{\tilde{\mathbf{p}}^2}{\sigma_{\tilde{\mathbf{p}}}^5}
%\to \frac{3}{d-1}\int \frac{d^{d-1}\tilde{\mathbf{p}}}{(2\pi)^{d-1}}\,\frac{\tilde{\mathbf{p}}^2}{\sigma_\mathbf{p}^5}
%= \frac{4}{(4\pi)^{d/2}}\Gamma\left(\frac{4-d}{2}\right)M^{d-4}(t).
%\eeq
%Hence the second term reads
%\beq
e^2\mathbf{J}_{\rm Q2} &  = & -
 \frac{e^2}{48\pi^2\hbar}\Gamma\left(\frac{4-D}{2}\right)\left(\frac{M^2(t)}{4\pi\mu^2}\right)^{(D-4)/2}
 \nonumber \\
 & = & \left( Z_3-1  + \frac{e^2}{48\pi^2\hbar}\log\frac{M^2(t)}{m^2}\right)\ddot{\mathbf{V}}, \label{Q2}
\eeqa
where we have used (see, e.g.~Ref.~\cite{Itzykson})
\beqa
Z_3 & = & 1- \frac{e^2}{48\pi^2\hbar}\Gamma\left(\frac{4-D}{2}\right)\left( \frac{m^2}{4\pi\mu^2}\right)^{(D-4)/2}\nonumber \\
& = & 1 - \frac{e^2}{48\pi^2\hbar}\left(\frac{1}{\varepsilon} - \gamma
- \log \frac{m^2}{4\pi\mu^2}\right).
\eeqa
Hence
\beqa
e^2\Delta \mathbf{J} & = & e^2\mathbf{J}_{\rm Q} - (Z_3-1)\ddot{\mathbf{V}}\nonumber \\
& = & e^2\mathbf{J}_{\rm Q1} + \frac{e^2}{48\pi^2\hbar}\ddot{\mathbf{V}}\log\frac{M^2(t)}{m^2}. \label{Qcorrect}
\eeqa
%where $\mathbf{J}_{\rm Q1}$ is given by Eq.~(\ref{Q1}).
%where we have introduced the renormalization scale $\mu$ in the usual manner.
%Now the $Z_3$ renormalization constant in scalar QED is given by (see Itzykson and Zuber, possibly)
%The original Lagrangian reads
%\beq
%{\cal L} = - \frac{1}{4}F_{\mu\nu}^B F^{B\,\mu\nu} - e_B A^B_\mu J^{{\rm c}\,\mu} + {\rm scalar\, term},
%\eeq
%where $F^B_{\mu\nu} = \partial_\mu A_\nu^B - \partial_\nu A^B_\mu$, which are {\rm bare} fields.  In terms of the renormalized %fields we have
%\beq
%{\cal L} = - \frac{Z_3}{4}F_{\mu\nu}^R F^{R\,\mu\nu} - e_R A^R_\mu J^{{\rm c}\,\mu} + {\rm scalar\,term}.
%\label{renLag}
%\eeq
%Note that $e_RA^R_\mu = e_B A^B_\mu$ because $A_\mu^B = Z_3^{1/2}A_\mu^R$ and $e_R = Z_3^{1/2}e_B$.  Then the field equation for the (renormalized) electromagnetic field including the vacuum-current contribution is
%\beq
%Z_3 \nabla_\nu F^{\nu\mu} = e_R (J^{{\rm c}\,\mu} + J^{{\rm VAC}\,\mu}).
%\eeq
%If we define the potential $W^\mu$ by $W^\mu = e_R A^\mu$, then 
%\beq
%Z_3 \nabla_\nu W^{\nu \mu} = e_R^2 (J^{{\rm c}\,\mu} + J^{{\rm VAC}\,\mu}).
%\eeq
%With the assumption that only the space components are nonzero and that they are only $t$-dependent, we have
%\beq
%Z_3 \ddot{\mathbf{W}} = e_R^2(\mathbf{J}^{{\rm c}} + \mathbf{J}^{{\rm VAC}}).  \label{Qeq}
%\eeq
%To lowest order, $Z_3=1$ and $\mathbf{J}^{{\rm VAC}} = 0$ so $\mathbf{W} = \mathbf{V}$ (by definition) and
%$\ddot{\mathbf{V}} = e_R^2\mathbf{J}^{{\rm c}}$.
Thus, Maxwell's equations $\ddot{\mathbf{V}} = e^2\mathbf{J}_{\rm C}$ has quantum corrections at one loop at order $\hbar^{-1}$, and the corrected equations can be written as follows:
\beq
\ddot{\mathbf{V}} = e^2\left[ 1 + \frac{e^2}{48\pi^2\hbar}\log\frac{M^2(t)}{m^2}\right]\mathbf{J}_{\rm C}
+ e^2\mathbf{J}_{\rm Q1}.
\eeq

%
%However, if the time-dependent mass arises in the effective flat-space theory from a time-dependent conformal factor, then the %quantum correction at this order in $\hbar$ vanishes again. We first find
%\beq
%\mathbf{J}^{{\rm c}} + \mathbf{J}^{{\rm VAC2}} = Z_3(t)\mathbf{J}^{{\rm c}}, \label{36}
%\eeq
%where $Z_3(t)$ is obtained from $Z_3$ by replacing $m^2$ by $M^2(t)$.   This means that the field equation now becomes
%\beq
%\ddot{\mathbf{W}} = e^2\left( 1 + \frac{e^2}{48\pi^2\hbar}\log\frac{M^2(t)}{m^2}\right)\mathbf{J}^{\rm c} 
%+ e^2\mathbf{J}^{{\rm VAC}\,1}.
%\eeq
%Thus, the field equation for the classical field for a given input classical current $\mathbf{J}^{\rm c}$ is modified, and as 
%result the charged field is affected by this.

\subsection{One-loop correction to the time-dependent mass}\label{subsection2}

This subsection closely follows Appendix A of Ref.~\cite{HiguchiMartin06a}.
The interaction Hamiltonian density contributing to the forward-scattering amplitude can be written as
\beq
{\cal H}_I(x) = eJ^\mu A_\mu + \frac{e^2}{\hbar^2}\sum_{i=1}^3 A_i A_i :\varphi^\dagger \varphi:
- \frac{\delta M^2(t)}{\hbar^2}:\varphi^\dagger\varphi:, \label{Hamilton}
\eeq
where the current $J^\mu$ is given by Eq.~(\ref{currentJ}).  We let $\delta M^2(t) = (\delta m^2/m^2)M^2(t)$, where $\delta m^2$ is set to the standard value in the on-shell renormalization.  Using the dimensional regularization in the standard covariant perturbation theory, one has (see, e.g.~Ref.~\cite{Itzykson})
\beqa
\delta m^2 & = & 
\frac{e^2\mu^{4-D}}{\hbar}\int \frac{d^Dq}{(2\pi)^Di}\left\{
\frac{(p+q)^2}{\left[q^2-m^2+i\epsilon\right]\left[(p-q)^2+i\epsilon\right]}
- \frac{4}{\left[(p-q)^2+i\epsilon\right]}\right\} \nonumber \\
& = &
-\frac{3e^2m^2}{(4\pi)^2\hbar} \left(\frac{1}{\eps} - \gamma +\frac{7}{3}- \log\frac{m^2}{4\pi\mu^2}\right).\label{delta2}
\eeqa
It is convenient for later purposes to perform the $q_0$ integral with the following result:
\beqa
\delta m^2 & = & 
\frac{e^2\mu^{4-D}}{\hbar}\int \frac{d^{D-1}\mathbf{q}}{(2\pi)^{D-1}}\left\{
- \frac{(\mathbf{p}+\mathbf{q})^2}{4Kq_0}\left[ \frac{1}{q_0+K-p_0}+\frac{1}{K+p_0+q_0}\right]\right.\nonumber \\
&& \,\,\,\,\,\,\,\,\,\,\,\,\,\,\,\,\,\,\,\left.
+ \frac{3}{2K} + \frac{1}{4Kq_0}\left[
\frac{(p_0-q_0)^2}{q_0+K+p_0} + \frac{(p_0+q_0)^2}{q_0+K-p_0}\right]\right\}, \label{delta1}
\eeqa
with $\mathbf{K} \equiv \mathbf{p}-\mathbf{q}$ and $K\equiv\|\mathbf{K}\|$, where $q_0$ is now defined to be $\sqrt{\mathbf{q}^2+m^2}$.  

\begin{figure}
\begin{center}
\includegraphics{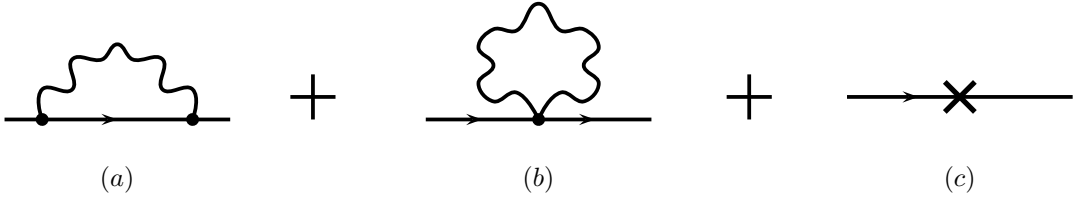}
\caption{The Feynman diagrams at order $e^2$ contributing to the correction to the mass term.}
\label{one-loopdiags1}
\end{center}
\end{figure}

The standard time-dependent perturbation theory to second order yields
\beq
{\cal F}(\mathbf{p})\langle \mathbf{p'}|\mathbf{p}\rangle
 =  -\int d^4x \langle \mathbf{p}'|\mathcal{H}(x)|\mathbf{p}\rangle
+ \frac{i}{2\hbar}T\left[\int d^4 x \int d^4 x'\,
\langle \mathbf{p}'|\mathcal{H}(x')\mathcal{H}(x)|\mathbf{p}\rangle\right]. \label{perturb}
\eeq
The third and second terms in the interaction Hamiltonian density
(\ref{Hamilton}) contribute to the forward-scattering amplitude through the first term on the right-hand side of Eq.~(\ref{perturb}).  They correspond to diagrams $(c)$ and $(b)$, respectively, in Fig.~\ref{one-loopdiags1} and their contribution to the forward-scattering amplitude is
\beqa
{\cal F}_c(\mathbf{p}) & = & - \frac{\delta m^2}{\hbar m^2}\int dt \,M^2(t)\frac{|\psi_\mathbf{p}(t)|^2}{2\sigma_\mathbf{p}(t)},\\
{\cal F}_b(\mathbf{p}) & = & - \frac{3e^2}{\hbar}\int dt\frac{|\psi_\mathbf{p}(t)|^2}{2\sigma_\mathbf{p}(t)} \int \frac{d^3\mathbf{q}}{2K(2\pi)^3}.
\label{1}
\eeqa
The first term in the interaction Hamiltonian density (\ref{Hamilton}) contributes through the second term in 
Eq.~(\ref{perturb}) with the corresponding Feynman diagram $(a)$ in Fig.~\ref{one-loopdiags1}.  It is, with the notation $x_1^\mu=(t_1,\mathbf{x}_1)$, and similarly for $x_2^\mu$,
\beq
{\cal F}_a(\mathbf{p})\langle\mathbf{p}'|\mathbf{p}\rangle 
= -ie^2 \int \frac{d^3\mathbf{k}}{2k(2\pi)^3} 
\int d^4x_1 d^4x_2 \theta(t_1-t_2)\langle \mathbf{p}'|J_\mu(x_1)J^\mu(x_2)|\mathbf{p}\rangle
e^{-ik\cdot(x_1-x_2)}.
\eeq
{}From this we obtain
\beqa\label{eq:<-D-->}
{\cal F}_a(\mathbf{p}) & = & 
\frac{ie^2}{2\hbar^2} \int \frac{d^3 \mathbf{q}}{2(2\pi)^3} \nonumber \\
&& \times 
\frac{1}{2K} \int dt_1 dt_2 \left\{\theta(t_1-t_2)[\phi_{\vect{p}}^*(t_1) \phi_{\vect{p}}(t_2) \overleftrightarrow{\mathcal{D}}_{\!\!1}(t_1,t_2,\vect{p},\vect{q}) \phi_{\vect{q}}(t_1) \phi_{\vect{q}}^*(t_2)] e^{-iK(t_1-t_2)/\hbar}\right.\nonumber \\
&&	\left.+ \theta(t_2-t_1) [\phi_{\vect{p}}^*(t_1) \phi_{\vect{p}}(t_2) \overleftrightarrow{\mathcal{D}}_{\!\!1}(t_1,t_2,\vect{p},\vect{q}) \phi_{\vect{q}}(t_2) \phi_{\vect{q}}^*(t_1)] e^{iK(t_1-t_2)/\hbar}\right\},
\eeqa
where $\mathbf{K} \equiv \mathbf{p}-\mathbf{q}$, $K\equiv \|\mathbf{K}\|$, and where
\begin{equation}
\overleftrightarrow{\mathcal{D}}_{\!\!1}(t_1,t_2,\vect{p},\vect{q}) \equiv -\hbar^2 \overleftrightarrow{\partial}_{\!\!t_1} \overleftrightarrow{\partial}_{\!\!t_2} + [\vect{p} + \vect{q}-2\mathbf{V}(t_1)] \cdot [\vect{p} + \vect{q}-2\mathbf{V}(t_2)].
\label{eq:<D>}
\end{equation}
(Note that the function $\phi_\mathbf{p}(t)$ differs from that in Ref.~\cite{HiguchiMartin06a} 
by a factor of $\sqrt{p_0}$.)
The $\hbar$ expansion can be constructed in exactly the same way as in the case where the mass term is constant.
Thus, we change the integration variables as $t_1 = t-\hbar\eta/2$ and $t_2=t+\hbar\eta/2$.  
Then, the expansion in $\hbar$ to lowest nontrivial order gives
\beq
{\cal F}_a(\mathbf{p}) = \frac{ie^2}{\hbar}\int \frac{d^3\mathbf{q}}{(2\pi)^3}
\frac{1}{2K}\int dt\left[G_{-}(\mathbf{p},\mathbf{q},t) + G_{+}(\mathbf{p},\mathbf{q},t)\right],
\label{Fa}
\eeq
where
\beqa
G_{\pm}(\mathbf{p},\mathbf{q},t)
& = & \pm \int_0^\infty d\eta \left[f_\pm(\mathbf{p},\mathbf{q},t)+O(\hbar^2)\right] \nonumber \\
&& \,\,\,\,\,\,\,\,\times \exp\left\{\mp i\int_{-\eta/2}^{\eta/2}d\zeta
\left[\pm\sigma_\mathbf{p}(t+\hbar\zeta)+\sigma_\mathbf{q}(t+\hbar\zeta)+K\right]\right\},\label{Gpm}\\
f_\pm(\mathbf{p},\mathbf{q},t) & = & \frac{1}{4\sigma_\mathbf{p}(t)\sigma_\mathbf{q}(t)}\left\{
-\left[\sigma_\mathbf{p}(t)\mp \sigma_\mathbf{q}(t)\right]^2 + \left[
\mathbf{p}+\mathbf{q}-2\mathbf{V}(t)\right]^2\right\}.
\eeqa
Now, note that
\beq
\exp\left\{\mp i\int_{-\eta/2}^{\eta/2}d\zeta
\left[\pm\sigma_\mathbf{p}(t+\hbar\zeta)+\sigma_\mathbf{q}(t+\hbar\zeta)+K\right]\right\}
= \exp \left\{ \mp i\left[\pm \sigma_\mathbf{p}(t) + \sigma_\mathbf{q}(t) + K\right]\eta\right\}
+ O(\hbar^2).
\eeq
Thus, we can integrate over $\eta$ in Eq.~(\ref{Gpm}) if we neglect the terms of 
higher order in $\hbar$, inserting a suitable infrared 
cutoff factor to regularize the integral for $\eta\to \infty$ as
\beq
\int_0^\infty d\eta\, \exp \left\{ \mp i\left[\pm \sigma_\mathbf{p}(t) + \sigma_\mathbf{q}(t) + K\right]\eta\right\}
= \frac{\mp i}{\pm \sigma_\mathbf{p}(t)+\sigma_\mathbf{q}(t)+K}.
\eeq
We define $\tilde{\mathbf{q}} \equiv \mathbf{q}-\mathbf{V}$, $\tilde{\mathbf{p}} \equiv \mathbf{p}-\mathbf{V}$,
$\tilde{q}_0 \equiv \sigma_\mathbf{q}(t)=\sqrt{\tilde{\mathbf{q}}^2 + M^2(t)}$ and 
$\tilde{p}_0 \equiv \sigma_\mathbf{p}(t) = \sqrt{\tilde{\mathbf{p}}^2 + M^2(t)}$, and change the variables of integration in Eq.~(\ref{Fa}) from $\mathbf{q}$ to $\tilde{\mathbf{q}}$.
As in the case with time-independent mass term, 
one encounters infrared divergences in higher-order terms in the $\hbar$ expansion,
but they do not contribute to the real part of the forward-scattering amplitude.  Thus, we obtain up to terms which vanish in the limit $\hbar \to 0$,
\beq
{\rm Re}\left[{\cal F}_a(\mathbf{p}) + {\cal F}_b(\mathbf{p})\right]
= -\int \frac{dt}{\sigma_\mathbf{p}(t)} \Delta M^2(t) + O(\hbar),
\eeq
where $\Delta M^2(t)$ is given by replacing $m^2$ by $M^2(t)$ in Eq.~(\ref{delta1}). (This result is not at all surprising because our semi-classical approximation boils down to evaluating diagram $(a)$ in Fig.~\ref{one-loopdiags1} with $M^2(t)$ treated as if it were time-independent in non-covariant perturbation theory.)  That is,
\beqa
\Delta M^2(t) & = & 
\frac{e^2\mu^{4-D}}{\hbar}\int \frac{d^{D-1}\mathbf{q}}{(2\pi)^{D-1}}\left\{
- \frac{(\tilde{\mathbf{p}}+\tilde{\mathbf{q}})^2}{4K\tilde{q}_0}\left[ \frac{1}{\tilde{q}_0+K-\tilde{p}_0}+\frac{1}{K+\tilde{p}_0+\tilde{q}_0}\right]\right.\nonumber \\
&& \,\,\,\,\,\,\,\,\,\,\,\,\,\,\,\,\,\,\,\left.
+ \frac{3}{2K} + \frac{1}{4K\tilde{q}_0}\left[
\frac{(\tilde{p}_0-\tilde{q}_0)^2}{\tilde{q}_0+K+\tilde{p}_0} 
+ \frac{(\tilde{p}_0+\tilde{q}_0)^2}{\tilde{q}_0+K-\tilde{p}_0}\right]\right\} \label{mudep} \nonumber \\
& = & -\frac{3e^2}{16\pi^2\hbar}M^2(t)\left( \frac{1}{\varepsilon} - \gamma + \frac{7}{3}
- \log \frac{M^2(t)}{4\pi^2\mu^2}\right),\label{M2}
\eeqa
where the terms which tend to zero as $\hbar \to 0$ have been dropped.

The correction (\ref{M2}) to the mass 
is not quite canceled out by the mass counterterm because of its logarithmic dependence on $M(t)$. 
Thus, the net correction to the squared mass is
\beq
\Delta M^2(t) -\delta M^2(t)
 =  \frac{3e^2}{16\pi^2\hbar}M^2(t) \log \frac{M^2(t)}{m^2}.  \label{subtraction}
\eeq
This correction will contribute to the forward-scattering amplitude as
\beq
{\rm Re}\left[{\cal F}_a(\mathbf{p}) + {\cal F}_b(\mathbf{p})+{\cal F}_c(\mathbf{p})\right]
= -\int \frac{dt}{\sigma_\mathbf{p}(t)} \left[\Delta M^2(t)-\delta M^2(t)\right] + O(\hbar),
\eeq
thus affecting the motion of the charged particle through Eq.~(\ref{loopPS}) 
at order $\hbar^{-1}$, i.e. \emph{at lower order} in the semi-classical approximation than the ALD force, which is of order $\hbar^0$.

\section{Vanishing one-loop corrections in Model B}\label{sec4}

As we have seen, the charged scalar field in conformally flat spacetime with the conformal factor depending only on time (Model B) is classically equivalent to that with a time-dependent mass in flat spacetime (Model A). However, this equivalence breaks down at one loop.  We show that the one-loop corrections we found for Model A in the previous section are canceled in Model B in this section.

In the dimensional regularization the conformal transformation $\phi = \Omega\varphi$ 
needs to be modified to $\phi = \Omega^{(2-D)/2}\varphi$.  There is no need to rescale $A_\mu$.  Then, the classical Lagrangian (\ref{eq:Lag}) is transformed to
\beq
\mathcal{L}  =  - \frac{1}{4}\Omega^{D-4}F_{\mu\nu}F^{\mu\nu} -\frac{1}{2}\Omega^{4-D}\left[\partial_\nu(\Omega^{D-4}A^\nu)\right]^2
+ ({\cal D}_\mu \varphi)^* {\cal D}^\mu\varphi - \frac{M^2_c}{\hbar^2}\varphi^*\varphi,
\eeq
with ${\cal D}_\mu = \partial_\mu + iV_\mu/\hbar + i(e/\hbar)A_\mu$, where indices are raised and lowered by the flat metric $\eta_{\mu\nu}$.   We have defined
\beq
M^2_c \equiv m^2\Omega^2 + \left[\frac{D-2}{2}-2(D-1)\xi\right]\hbar^2\left(
\partial_\mu\partial^\mu\log\Omega + \frac{D-2}{2}\partial_\mu\log\Omega\partial^\mu\log\Omega\right).  \label{dimconf}
\eeq
We also have inserted a gauge-fixing term which would correspond to the Feynman gauge if $\Omega(t)=1$.

First we show that the relation between the background field $V^\mu$ and the classical current $J_{\rm C}^\mu$ is unchanged at one loop 
in the limit $\hbar \to 0$.  The correction to the current, Eq.~(\ref{Z3-1}), for Model A is changed here to
\beq
e^2\Delta J^\mu = e^2J_{\rm Q}^\mu - (Z_3-1)\partial_\nu \left[\Omega^{D-4}(\partial^\nu V^\mu -
\partial^\mu V^\nu)\right].
\eeq
Since the currents and the field $V_\mu$ depend only on time and $J^0_Q=V_0=0$, we have
\beq
e^2 \Delta \mathbf{J} = e^2\mathbf{J}_{\rm Q1} + e^2\mathbf{J}_{\rm Q2} 
-(Z_3-1)
\frac{d\ }{dt}\left(\Omega^{-2\varepsilon}\dot{\mathbf{V}}\right).
\eeq
Since $\varepsilon(Z_3-1) \to  -e^2/(48\pi^2\hbar)$ as $\varepsilon\to 0$, we find
\beqa
e^2\Delta \mathbf{J} & = & e^2\mathbf{J}_{\rm Q1} + e^2\mathbf{J}_{\rm Q2}
- \frac{e^2}{48\pi^2\hbar}\left(\frac{d\ }{dt}\log\Omega^2\right)\dot{\mathbf{V}}
- \left(Z_3 - 1 + \frac{e^2}{48\pi^2\hbar}\log\Omega^2\right)\ddot{\mathbf{V}},
\eeqa
where $\mathbf{J}_{\rm Q1}$ and $\mathbf{J}_{\rm Q2}$ are given by Eqs.~(\ref{Q1}) and (\ref{Q2}) with
$M^2(t)$ replaced by $M_c^2(t)$.
We can replace $M_c^2(t)$ by $m^2\Omega^2(t)$ in the limit $\hbar \to 0$ because the difference between these quantities is of order $\hbar^2$ [see Eq.~(\ref{dimconf})].  Thus,
$\Delta \mathbf{J} = 0$, and hence Maxwell's equations 
do not get any corrections at order $e^2$ in the limit $\hbar \to 0$.

\begin{figure}
\begin{center}
\includegraphics{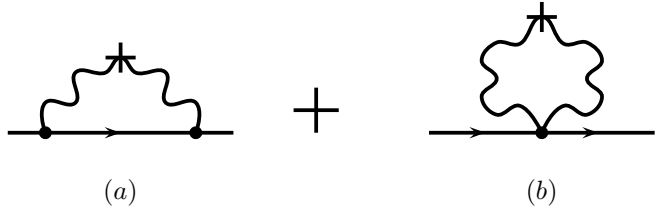}
\caption{The Feynman diagrams contributing to the correction due to the quadratic interaction term.  The cross indicates the ``interaction term'' $-[(4-D)/2]Q_{\mu\nu}A^{\mu}A^{\nu}$.}
\label{UVdivergent}
\end{center}
\end{figure}

Next we show that there is no correction to the time-dependent mass term in the limit $\hbar \to 0$.
We first need to discuss the modification in the free electromagnetic field equations due to the change in dimensions in more detail.
{}From the Lagrangian describing the free electromagnetic field,
\beq
\mathcal{L}_{\rm EM} \equiv - \frac{1}{4}\Omega^{D-4}F_{\mu\nu}F^{\mu\nu} -\frac{1}{2}\Omega^{4-D}\left[\partial_\nu(\Omega^{D-4}A^\nu)\right]^2,
\eeq
we find the following field equation:
\beq
\partial_\nu(\Omega^{D-4}\partial^\nu A^\mu) + (D-4)\Omega^{D-4}(\partial^\mu\partial_\nu \log\Omega)A^\nu = 0.
\eeq
This equation can be written
\beq
\Omega^{(D-4)/2}\partial_\nu \partial^\nu \left[ \Omega^{(D-4)/2}A_\mu\right] + \frac{4-D}{2}Q_{\mu\nu}A^\nu = 0,
\eeq
where
\beq
Q_{\mu\nu} \equiv \Omega^{D-4}\left[ 2\partial_\mu\partial_\nu \log\Omega
- \eta_{\mu\nu}\partial_\alpha\partial^\alpha\log\Omega
+\frac{4-D}{2} \eta_{\mu\nu} \partial_\alpha\log\Omega \partial^\alpha\log\Omega\right].
\eeq
Thus, if we write
\beq
{\cal L}_{\rm EM} = {\cal L}_{\rm free} -\frac{4-D}{4}Q_{\mu\nu} A^\mu A^\nu, \label{artificial}
\eeq
where
\beq
{\cal L}_{\rm free} = - \frac{1}{4}\Omega^{D-4}F_{\mu\nu}F^{\mu\nu} - \frac{1}{2}\Omega^{4-D}
\left[\partial_\nu (\Omega^{D-4}A^\nu)\right]^2 + \frac{4-D}{4}Q_{\mu\nu}A^\mu A^\nu,
\eeq
then the Euler-Lagrange equation from ${\cal L}_{\rm free}$ is
$\partial_\nu \partial^\nu (\Omega^{(D-4)/2}A^\mu) = 0$.
Hence, by regarding the second term in Eq.~(\ref{artificial}) as an interaction term, we can expand $A_\mu$ in the interaction picture, i.e.\ as a free field satisfying this Euler-Lagrange equation, as
\beq
A_\mu(x) = \Omega^{(4-D)/2}(t)\int \frac{d^3\mathbf{k}}{2k(2\pi)^3}
\left[a_\mu(\mathbf{k})e^{-ik\cdot x} + a_\mu^\dagger(\mathbf{k})e^{ik\cdot x}\right]. \label{Amumode}
\eeq
One can readily verify that the canonical quantization leads to the standard commutation relations (\ref{commutators}).
The ``interaction term'' $-[(4-D)/4]Q_{\mu\nu}A^\mu A^\nu$ in Eq.~(\ref{artificial}) contributes 
only in the ultraviolet divergent diagrams, which are given by Fig.~\ref{UVdivergent}, because of the factor $4-D = 2\varepsilon$. 
One expects that these diagrams will generate a term proportional to
$e^2\hbar {Q^\alpha}_\alpha(t):\varphi^\dagger \varphi:$ in the one-loop effective Lagrangian.  
We show that this is indeed the case in Appendix A.
Thus, the electromagnetic Lagrangian $\mathcal{L}_{\rm EM}$ may be changed to ${\cal L}_{\rm free}$, which leads to
the mode expansion (\ref{Amumode}), and 
we only need to incorporate the factor $\Omega^{(4-D)/2}$ in this mode expansion to
to adapt the calculations in Sec.~\ref{subsection2} to Model B.
Thus, in Eq.~(\ref{1}) the integrand is now multiplied by $\Omega^{4-D}(t)$ and in Eq.~(\ref{eq:<-D-->}) the integrand is multiplied by $\Omega^{(4-D)/2}(t_1)\Omega^{(4-D)/2}(t_2)$.  In the latter equation the change of variables
$t_1 = t-\hbar\eta/2$ and $t_2 = t+\hbar\eta/2$ leads to
\beq
\Omega^{(4-D)/2}(t_1)\Omega^{(4-D)/2}(t_2)
= \Omega^{4-D}(t) + O(\varepsilon^2\hbar^2\eta^2).
\eeq
Terms of order $\varepsilon^2$ do not contribute at one loop because there are only simple poles in $\varepsilon$. 
Thus, the integrand for $\mathcal{F}_b(\mathbf{p})$ in Eq.~(\ref{1}) and $\mathcal{F}_a(\mathbf{p})$ in
Eq.~(\ref{Fa}) are both multiplied by $\Omega^{4-D}(t)$.
Hence, the factor $\mu^{4-D}$ in Eq.~(\ref{mudep}) is (put inside the integral sign and) replaced by $(\mu\Omega)^{4-D}$, and the function $\log[M^2(t)/\mu^2]$ is changed to $\log[M^2_c(t)/\mu^2\Omega^2] = \log(m^2/\mu^2) + O(\hbar^2)$.  Thus, in Eq.~(\ref{subtraction}) 
$\Delta M^2(t)$ is changed to $\delta m^2\Omega^2$ 
in the limit $\hbar\to 0$.  Hence the correction to the time-dependent mass, $\Delta M^2_c(t)-\delta M^2(t)$, and the corresponding contribution to the position shift $\delta_{\rm loop}x^i$ in Eq.~(\ref{loopPS}) vanish
in the limit $\hbar \to 0$.

\section{Concluding remarks and outlook}\label{sec5}

We investigated in this paper the physics of a charged scalar particle moving in conformally flat spacetime in quantum electrodynamics. The conformal factor was assumed to depend only on time.  First we showed, using the conformal transformation to flat spacetime, that the classical radiation-reaction force of DeWitt, Brehme and Hobbs is reproduced by the one-photon emission process in quantum electrodynamics in the limit $\hbar\to 0$, closely following the work of Higuchi and Martin.  Then we investigated the one-loop corrections in the WKB approximation for the charged scalar field in flat spacetime with time-dependent mass (Model A) and that in conformally flat spacetime with the conformal factor depending only on time 
(Model B), which is classically equivalent to Model A.  We found that the quantum corrections in Model A do not vanish in the limit $\hbar \to 0$ but those in Model B, which is of more interest in the context of this paper, vanish.  This discrepancy is due to the fact that these two models are not equivalent quantum mechanically.

It will be interesting to extend our observations in this paper and investigate the connection between the one-photon emission process and the non-tail terms in the classical DeWitt-Brehme-Hobbs radiation-reaction formula in general spacetime.  It will also be interesting to analyze the one-loop QED corrections in general spacetime to determine whether there can be any large quantum corrections in radiation processes.  We cannot rule out this possibility because the one-loop corrections are naturally of order $\hbar^{-1}$ as we have seen in Model A.  In this context it is worth pointing out that the tail terms in the DeWitt-Brehme-Hobbs formula and the MiSaTaQuWa formula for the gravitational radiation reaction~\cite{MiSaTa,QuinnWald} are expected to come from one-loop diagrams in quantum field theory because the tail terms represent self-interaction of matter fields through the electromagnetic and gravitational fields.

\acknowledgments

One of the authors (A.H.) thanks the Astro-Particle Theory and Cosmology Group and the Department of Applied Mathematics at University of Sheffield, where part of this work was carried out, for kind hospitality.

\appendix

\section{Demonstration that the ``interaction term'' in Eq.~(\ref{artificial}) is negligible}

In general let $\hat{Q}_{\mu\nu}(k)$ be the Fourier transform of $Q_{\mu\nu}(x)$.  That is
\beq
Q_{\mu\nu}(x) = \int \frac{d^4k}{(2\pi)^4} \hat{Q}_{\mu\nu}(k)e^{-ik\cdot x}.
\eeq
The Feynman diagrams in Fig.~\ref{UVdivergent} are evaluated as
\begin{eqnarray}
\hat{\Sigma}(p,k) & = & e^2\mu^{4-D}\varepsilon
\int \frac{d^Dq}{(2\pi)^Di}\left\{ \frac{(p^\mu+q^\mu)(p^\nu + k^\nu + q^\nu)\hat{Q}_{\mu\nu}(k)}{
(q^2-m^2+i\epsilon)\left[(p-q)^2+i\epsilon\right]\left[(p+k-q)^2+i\epsilon\right]} \right.\nonumber \\
&& \left.\ \ \  \ \ \ \ \ \ \ \ \ \ \ \ \ \ \ \ \ \ \ \ \ 
- \frac{\eta^{\mu\nu}\hat{Q}_{\mu\nu}(k)}{(q^2+i\epsilon)\left[(q-k)^2+i\epsilon\right]}\right\}.
\end{eqnarray}
(We have let $\hbar=1$ here.)
Both terms in this integral are only logarithmically divergent. Hence, to find the pole in $\varepsilon = (4-D)/2$ we may replace all factors of the form $q^2 + \cdots$ in the denominator by $q^2 - \lambda^2+i\epsilon$, where $\lambda$ is an arbitrary positive number.  Thus we have 
\beqa
\hat{\Sigma}(p,k) & = & \varepsilon e^2 \mu^{2\varepsilon}\int \frac{d^Dq}{(2\pi)^Di}\left[
\frac{q^\mu q^\nu \hat{Q}_{\mu\nu}(k)}{(q^2-\lambda^2 + i\epsilon)^3}
- \frac{{{\hat{Q}}^\alpha}_{\ \alpha}(k)}{(q^2-\lambda^2 + i\epsilon)^2}\right] \nonumber \\
& = & - \frac{3}{64\pi^2}e^2{{\hat{Q}}^\alpha}_{\ \alpha}(k).
\eeqa
Hence the contribution to the effective Lagrangian is
\beq
{\cal L}_{\rm eff} = - \frac{3e^2\hbar}{64\pi^2}\, {Q^\alpha}_\alpha(x):\varphi^\dagger(x)\varphi(x):,
\eeq
where we have inserted a factor of $\hbar$ by dimensional analysis.
Thus, the contribution of the extra mass-like term in Eq.~(\ref{artificial}) 
to the effective Lagrangian vanishes in the limit $\hbar \to 0$.

%%%%%%%%%%%%%%%%%%%%%%%%%%% References %%%%%%%%%%%%%%%%%%%%%%%%%%%%%

\end{document}